# PANDEMIC RECESSION AND HELICOPTER MONEY: VENICE, 1629-1631♠

Charles Goodhart♠♠ Donato Masciandaro♠♠♠ Stefano Ugolini ♠♠♠♠

December 4, 2021

Abstract

We analyse the money-financed fiscal stimulus implemented in Venice during the famine and plague of 1629-31, which was equivalent to a "net-worth helicopter money" strategy – a monetary expansion generating losses to the issuer. We argue that the strategy aimed at reconciling the need to subsidize inhabitants suffering from containment policies with the desire to prevent an increase in long-term government debt, but it generated much monetary instability and had to be quickly reversed. This episode highlights the redistributive implications of the design of macroeconomic policies and the role of political economy factors in determining such designs.

---

♠ The authors warmly thank Guido Alfani, Pierpaolo Benigno, Jordi Galì, Athanasios Orphanides, Luciano Pezzolo, Franco Spinelli, Mara Squicciarini, Carolyn Sissoko, Gianni Toniolo and Geoffrey Wood for their kind and useful comments and reactions on earlier drafts. Fabio Gatti provided excellent research assistance; his effort in finding out archival sources deserves special thanks. We are also grateful to Rui Pedro Esteves and to two anonymous referees for their help in improving our manuscript. The usual disclaimers apply.

♠♠ London School of Economics.

♠♠♠ Bocconi University.

♠♠♠♠ University of Toulouse (Sciences Po Toulouse and LEREPS).

Whenever economic conditions become really critical, the concept of *helicopter money* frequently re-emerges – and the recent Covid-19 crisis was no exception (Masciandaro 2020). The term "helicopter money" originates from Milton Friedman, who famously imagined (without endorsing it) a hypothetical situation in which freshly-issued central bank money would be randomly distributed to households.[1] Friedman's description, however, was characteristically short in detail on how such a monetary policy might concretely be implemented. As a result, while followers have agreed on the definition of helicopter money as "a money-financed fiscal stimulus […] that […] requires neither an increase in the stock of government debt nor higher taxes, current or future" (Galí 2020, p. 2), they have diverged in their views of the design of such a policy. On the one hand, most economists have interpreted helicopter money as a permanent increase in the liabilities of the central bank (see esp. Buiter 2014). As such a situation, which we might describe as "*monetary-base helicopter money*", has actually occurred relatively often in history, it represents a less radical option than it might appear at first sight. On the other hand, others have interpreted helicopter money as a permanent decrease in the assets (i.e., in the net worth) of the central bank (see esp. Masciandaro 2020). Such a situation, which we might describe as "*net-worth helicopter money*", is a much more radical option than the previous one, and appears to have occurred only on exceedingly rare occasions in history.

In order to better understand the difference between "monetary-base" and "net-worth helicopter money", imagine a stylized balance sheet of the public sector made of the separate balance sheets of the fiscal authority (the Treasury) and of the monetary authority (the central bank: see Figure 1). The fiscal authority's assets consist of Treasury deposits with the central bank (*TD*) and of other Treasury assets (*TA*); its liabilities consist of the Treasury's net worth (*TW*), of marketable government bonds (*TB*), and of direct (unmarketable) loans from the central bank (*TL*). The monetary authority's assets consist of its portfolio of marketable government bonds (*TB*),[2] of its bullion and foreign reserves (*BR*), and of its direct

[1] "Let us suppose now that one day a helicopter flies over this community and drops an additional $1,000 in bills from the sky, which is, of course, hastily collected by members of the community. Let us suppose further that everyone is convinced that this is a unique event which will never be repeated" (Friedman 1969, pp. 4-5).

[2] Note that contrary to Treasury deposits with the central bank (*TD*) and to central bank's direct loans to the Treasury (*TL*), the amount of marketable government bonds (*TB*) on the monetary authority's balance sheet will not necessarily be equal to the amount of the same item on the fiscal authority's balance sheet: as a matter of fact, a substantial amount of marketable government bonds will be typically held by the private sector.

(unmarketable) loans to the Treasury (*TL*); its liabilities consist of its net worth (*BW*),[3] of the Treasury's deposits (*TD*), and of the deposits of the private sector – i.e. the monetary base (*MB*).[4]

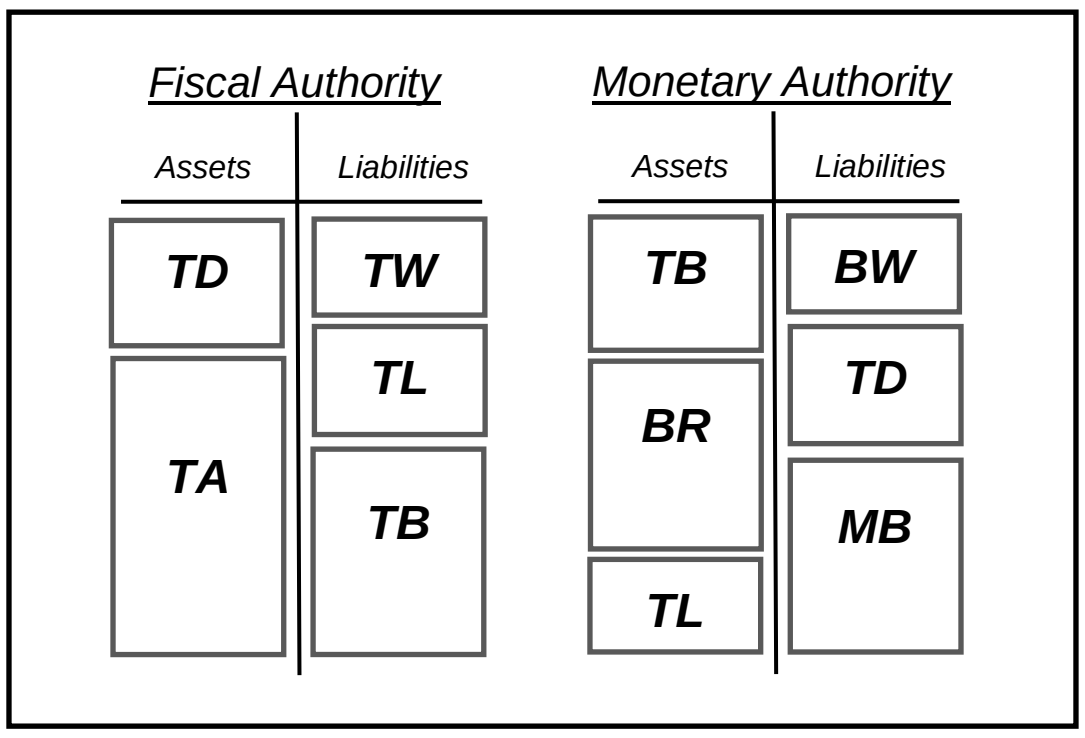

*Figure 1. Stylized balance sheets of the fiscal and monetary authorities.*

"Monetary-base helicopter money" occurs in two steps. First, the Treasury issues new bonds and these are all bought by the central bank (on either the primary or the secondary market), which allows for a temporary increase in Treasury deposits with the central bank (Figure 2.1). Then, the Treasury proceeds to spend this money, which ends up being held by the private sector: the monetary base hence increases. On the fiscal authority's balance sheet, the money-financed fiscal stimulus reduces the Treasury's assets without reducing any liability except the Treasury's net worth, and can therefore be regarded as a loss (Figure 2.2). On the whole, the difference between this "monetary-base helicopter money" strategy and a classical fiscal expansion is quantitative rather than qualitative, as the boundary between the two is defined exclusively by the size of the government debt monetization by the central bank.

[3] The monetary authority's net worth (*BW*) will include central bank equity when the latter formally exists. Note that central bank equity may be owned by the fiscal authority – in which case, it will be included in the Treasury's assets (*TA*) – but this will not always necessarily be the case.
[4] Note that the monetary authority's and fiscal authority's balance sheets might substantially differ in terms of their size.

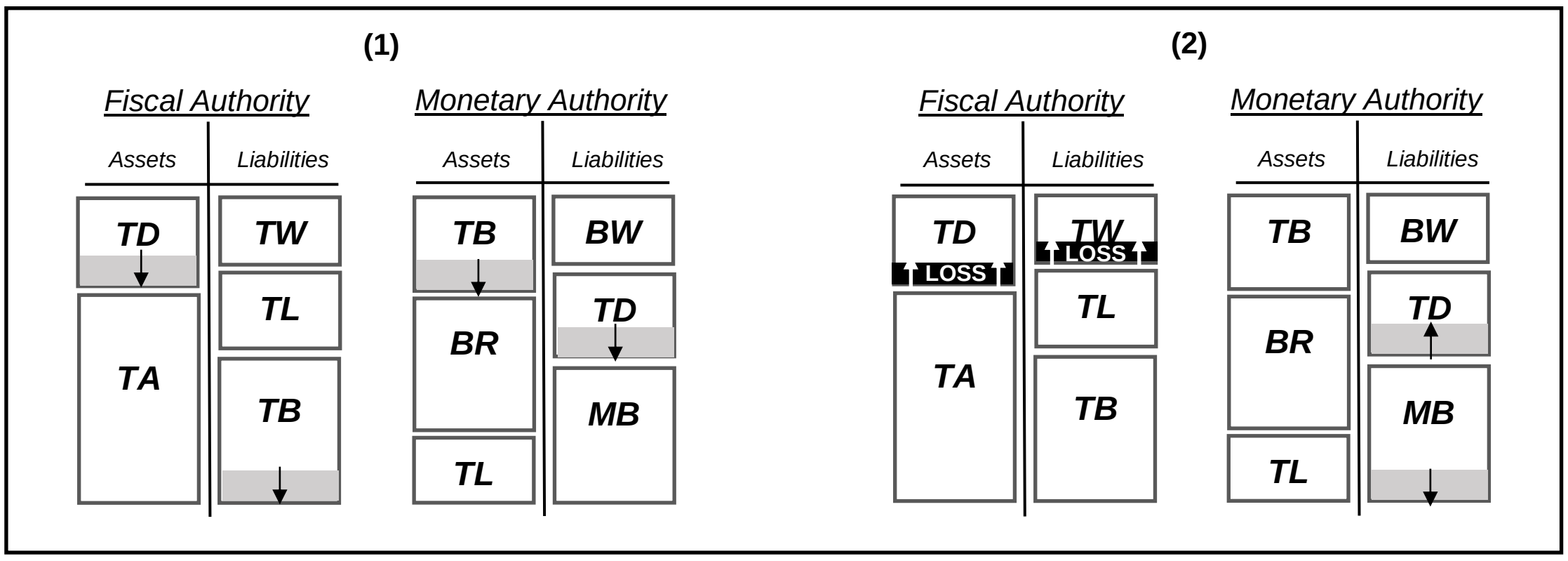

*Figure 2. Balance sheet effects of a "monetary-base helicopter money" strategy.*

By contrast, a "net-worth helicopter money" strategy is qualitatively different from a classical fiscal expansion. Here, the fiscal authority instructs the monetary authority to credit money on the private sector's accounts with it. This increase in the central bank's liabilities should be theoretically matched by an increase in its assets – viz., in its direct loans to the Treasury (Figure 3.1). In truth, however, these loans simply do *not* formally exist – which is economically equivalent to a Treasury default on its direct borrowings from central bank. As a result, the money-financed fiscal stimulus here reduces the *monetary* authority's (instead of the fiscal authority's) assets without reducing any of its liability except the net worth, and can therefore be regarded as a loss (Figure 3.2). From a long-term perspective, "net-worth helicopter money" is therefore irreversible for the monetary authority, whereas "monetary base helicopter money" might potentially be (partially or totally) reversed by selling marketable Treasury bonds to the private sector. The biggest difference between the two is however of short-term nature, as "net-worth helicopter money" fragilizes the solidity of monetary authority and directly reduces its margins for manoeuvre in the management of the value of issued money, thus rendering it less able to counteract an inflationary outburst.[5]

[5] The literature has generally only considered long-term implications, arguing that the only difference between irreversible "monetary base helicopter money" and "net-worth helicopter money" is the fact that the former affects the Treasury's expenditures while the latter affects its revenues (as the Treasury is the only recipient of the central bank's dividends, which are negatively impacted by the latter strategy: see esp. Buiter 2014). This conclusion is based on the assumption that the monetary authority is economically indistinguishable from the fiscal authority – which may not actually be always true, as it is the case today e.g. in the Eurozone, and as it was very often the case in the past. By contrast, the short-term implications of each strategy on monetary policy implementation have not been considered so far.

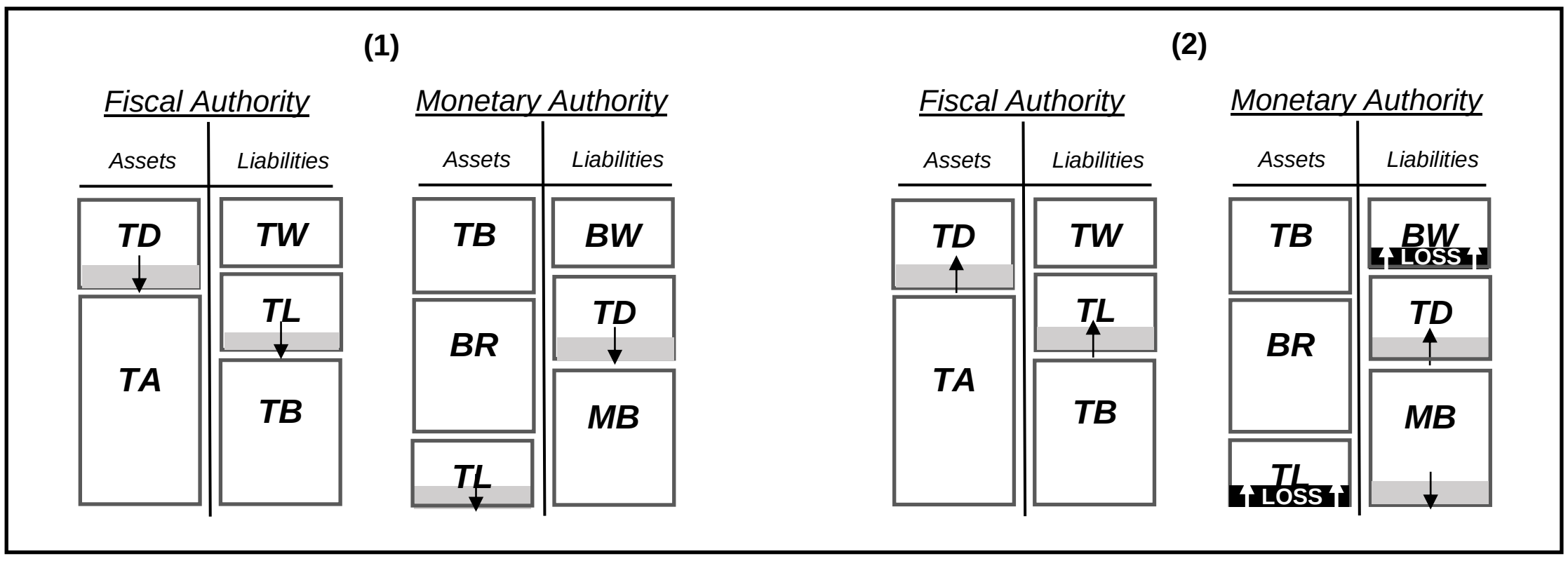

*Figure 3. Balance sheet effects of a "net-worth helicopter money" strategy.*

Theoretically, it is unclear whether one version of "helicopter money" is always necessarily superior to the other. Empirically, the implementation of "net-worth helicopter money" has been so rare in history that it has been practically impossible to analyse it. In this article, we study one rare historical episode in which "net-worth helicopter money" appears to have actually been implemented. This happened in the Republic of Venice in 1629-31 – interestingly, precisely in the context of a full-fledged pandemic which bears many resemblances with Covid-19. We describe the historical context in which fiscal monetization took place, explain why it can be considered as an example of "net-worth helicopter money", and analyse its consequences. In particular, we ask what were the political economy conditions which generated the choice of this very peculiar version of expansionary fiscal policy.

The remainder of this article is organised as follows. In Section I we briefly describe the Venetian monetary and fiscal systems during the first decades of the seventeenth century. In Section II we analyse the huge real economic shocks of 1629-1631, as well as the containment policies put in place by the Venetian government during this period. Section III focuses on the "net-worth helicopter money" strategy implemented in order to finance expenditure, and offers a political perspective to help explain why the Republic chose this radical option rather than more classical ones. Section IV concludes.

# I

The Republic of Venice was run for centuries by a small and cohesive élite of oligarchs ("patricians") highly invested in mercantile activities. Citizenship did not automatically guarantee political rights, but it gave complete access to local welfare, guaranteeing protection in times of crisis. Residing and working in Venice were not sufficient conditions to gain access to public services, which only citizenship did insure (Alfani and Di Tullio 2019, pp. 59-60). In fact, only the patricians had political rights, assisted by the citizens, who carried out bureaucratic tasks; ordinary inhabitants, neither patrician nor citizen, were excluded from any political or administrative participation (Finlay 1980). Yet, although formally excluded from any involvement in politics and institutions, inhabitants used civil rituals, crowd behaviour, and collective actions to influence patricians' decision making, especially in times of crisis (van Gelder 2018, p. 254; Judde de Larivière 2020, p. 72).

In Venice, the relative size of the public sector with respect to the domestic economy was extraordinarily high for coeval standards. In the real sector, the government, via the Grain Office and then the Fodder Office, was active in the grain market, in order to address and to stabilize the volatility in the supply and price of food, which was an extremely sensitive political issue. As for the financial sector, the financial district of Venice – the Rialto area – was entirely owned by the Republic, which rented to the private bankers the benches at which they used to operate in what was considered as a public concession. Private bankers' books were considered public records, the bank transfers being a legal way to discharge debt under Venetian law (Ugolini 2017, pp. 37-9). Yet the ideal that inspired the Venetian government was that, as far as possible, the State should not replace private initiative in markets (Dunbar 1892, p. 308). The Senate was the body that designed the Republic's economic policy, with its committees responsible for overseeing every aspect of production and distribution (Rapp 1976, p. 139). With the Republic being firmly controlled by an oligarchy of merchants, the government's goal was just to provide the services that were essential for business but too expensive or too risky to be provided by the businessmen themselves (Ugolini 2017, p. 38).

The Republic's tax system, featuring both "direct" and "indirect" taxes, was rather muddled and complicated. Reflecting the balance of power, the system had a marked "regressive" nature. From the second half of the sixteenth century until the 1620s, the Republic of Venice gradually increased per capita taxation, and this growth accelerated much more markedly in the aftermath of the 1630 plague (Alfani and Di Tullio 2019, pp. 24-33 and 147). The Republic had a long and generally good record with public debt management, which had been implemented since at least 1164. In early times, Venice had regularly resorted to forced loans from the patricians; by the seventeenth century a more market-friendly approach had prevailed thus increasingly attracting foreign investors, yet the government debt was still overwhelmingly held by the domestic upper class (Fratianni and Spinelli 2006, pp. 262-3; Alfani and Di Tullio 2019, p. 172).[6] The Republic issued both floating and funded (long-term) debt, though usually the role of floating debt was limited; in the course of the sixteenth century, the management of the funded debt had been entrusted to the State-owned Mint (Pezzolo 2003a). In tranquil years, the Republic generated substantial fiscal surpluses in order to repay all its debt (a goal which had been fully attained, for instance, in 1600), thus enhancing its creditworthiness (Sissoko 2002, p. 8; Fratianni and Spinelli 2006, p. 263; Alfani and Di Tullio 2019, p. 172). Therefore, at the beginning of the seventeenth century the state of public finance in Venice was good (Pezzolo 2003a, p. 103), and the government borrowed at an interest rate which was substantially lower than the average rate of return from trade (Sissoko 2007, p. 6).

By the seventeenth century, Venice had reached a degree of monetization unknown for centuries anywhere else. As usual in early-modern times, the official monetary anchor consisted of commodity money (coins) issued by the Mint (Al-Bawwab 2021); however, inhabitants commonly used cheques and bank transfers (even among the lower middle class), which allowed them to economize on coins (Mueller 1997, p. 24; Fratianni and Spinelli 2006, p. 271). This had been made possible by the development of the private deposit banks operating on the Rialto square, but by the end of the 16th century all such banks had gone bust and no serious private initiative was any longer available for running the business on the conditions imposed by regulation (Ugolini 2017, pp. 35-43). As a result, the State eventually took the initiative on itself, and on April 1587 Venice's first public bank (called "Banco della Piazza di Rialto") started its operations (Soresina 1889, p. 7; Roberds and Velde 2014, p. 19; Bindseil 2019, pp. 207-10).

In theory, the Banco di Rialto was supposed to represent a case of quasi-narrow banking, given that it was obliged by law to accept only deposits in coins, and cash was always to remain available at the request of depositors (Anonimo 1847, p. 564; Dunbar 1892, p. 321), defining tendentially a policy of 100 per cent reserves (Sissoko 2002, pp. 7 and 10); transfers had to be made simultaneously between creditors and debtors (Dunbar 1892, p. 321; Roberds and Velde 2014, p. 20). In practice, however, because the coins into which the Banco's liabilities were formally convertible had been withdrawn from circulation in 1588, bank liabilities were de facto inconvertible into the new circulating coins (Roberds and Velde 2014, pp. 21-2; Ugolini 2017, pp. 225-6). Coins were the main item on the asset side of the Banco's balance sheet, also private commercial debt was allowed to a certain extent (Ugolini 2017, p. 44). In 1593 Banco liabilities became legal tender (Anonimo 1847, p. 366; Roberds and Velde 2014, p. 21); by 1630, its deposits in represented 80% of the overall volume of exchange settlements in Venice (Sissoko 2002, p. 8; Roberds and Velde 2014, p. 21). During these years the two legal tenders – commodity and scriptural money – were imperfect substitutes, the conversion rate between the two being determined on the market (Dunbar 1892, p. 318; Inclimona 1913, p. 149; Fratianni and Spinelli 2006, p.271). The existence of a positive premium ("*agio*") for Banco money with respect to coins was almost a constant in the Venetian experience (Siboni 1892, p. 291; Magatti, 1914, pp. 285-9; Roberds and Velde 2014, p. 17). As a matter of fact, scriptural money at the time was safer for obvious reasons, and the quality of available commodity money was especially poor during the "bullion crisis" of the late sixteenth and early seventeenth century (Dunbar 1892, pp. 309, 321, and 330-1; Ugolini 2017, pp. 225-7).

[6] Despite the share of the Venetian funded debt held by foreigners constantly grew throughout the sixteenth and seventeenth centuries, as late as in 1641 up to six sevenths of this debt were still held by citizens (Alfani and Di Tullio 2019, p. 172).

In May 1619 the government created a new public bank – the Banco del Giro – with floating (short-term) public debt and coins on the asset side of its balance sheet and deposits on the liability side (Soresina 1889, p. 9, Siboni 1892, p. 288, Inclimona 1913 p.152, Roberds and Velde 2014, p.24; Bindseil 2019, pp. 215-7). In general, the State's creditors were likely to become floating debt holders using the transfer mechanism of the State bank. This sort of mechanism had been first introduced in the thirteenth century when the Grain Office and Salt Office had started providing transfers to their creditors, and also the Fodder Office had resorted to it from 1608 to 1614 (Pezzolo 2003a, p. 63; Roberds and Velde 2014, p. 24; Ugolini 2018, p. 6). While the Banco della Piazza di Rialto was a deposit bank, the Giro bank was a device to make the public debt easily transferable, turning it into a means of payment (Roberds and Velde 2014, p. 22), and "paying deposits at the call of the depositor, like the existing Banco di Rialto", with the possibility of deposit overdrawing, i.e. making loans (Dunbar 1892, p. 325). The account holders were floating debt holders; the Giro bank was allowed to accept deposits of private individuals only after the closure of the Banco della Piazza (Sissoko 2002, p. 11).

The functioning of the Giro bank was simple (Ugolini 2017, p. 43): the government opened accounts to merchants having credits to the Republic and to public officials as well (Soresina 1889 p.16, Inclimona 1913 p.1146), that could be converted into coins upon authorization; merchants and public officials became Republic's debt holders. The credit of one accountholder could always be freely transferred on demand to another accountholder, and the corresponding amount would continue to circulate until the final repayment to the last bearer cancelled it out (Soresina 1889, pp. 12 and 16). Banknotes were not issued (Siboni, 1892, p. 291). The Giro bank liabilities were legal tender for any payment greater than one hundred ducats, while its clearing activity was possible also for payments lower than one hundred ducats (Soresina 1889 p.12). Moreover, from July 1627 the account holders could pay import taxes using Giro bank transfers (Soresina 1889, p. 20; Siboni 1892, p. 294). The convertibility promise on Giro bank deposits was based on the fact that in the State Mint an amount of commodity money served as a fund to back the operations of the Giro (Dunbar 1892, p. 326), although backing was not 100 per cent. In fact, on June 1619 the Senate authorized on the one hand the creation of 150,000 ducats' worth of coin reserves earmarked at the Mint for the Giro, and on the other hand 500,000 ducats' worth of Giro balances to pay its creditors – i.e., the would-be Giro depositors (Soresina 1889 pp. 12-3; Roberds and Velde 2014, p. 23). Moreover, one part of the Mint's output was earmarked to repay Giro scriptural money using coins: the decree of foundation of the Banco actually ordered monthly transfers of 10,000 ducats from the Mint to the Giro for repayments, up to the limit of 50,000 ducats (Roberds and Velde 2014, p. 24; Soresina 1889, p. 15). On January 1620 the overall balance and the monthly transfers became respectively 700,000 ducats and 20,000 ducats; the monthly transfers eventually became 80,000 ducats in August 1625 (Soresina 1889, p. 17; Siboni 1892, p. 288; Roberds and Velde 2014, p. 24). The Giro balances were further increased on May 1621 – by 100,000 ducats – and on June 1621 – by 40,000 ducats (Soresina 1889, p. 18). Then, "as long as the monthly flow was sufficient to accommodate depositors' requests, the bank's liabilities remained convertible […]; the State […] adjusted the monthly flows of cash from the Mint to service the redemption requests" (Roberds and Velde 2014, p. 24).

All in all: from 1619 a duopolistic public banking system was born in Venice, where the liabilities of the two banks were treated as equivalent (Dunbar 1892, p. 324; Ugolini 2017, p. 44) and both granted the seizure exemption privilege, meaning that in no case did judicial courts have the power to seize their deposits (Soresina 1889, p. 8). Moreover, in their period of coexistence the two public banks were interconnected in some coin exchange operations; while the reciprocal clearing of their liabilities was forbidden, given the need to maintain separation between the two banks (Soresina 1889, pp. 9 and 13). The duopolistic setting ended in 1637, when the Banco della Piazza di Rialto was shut down, with the Banco del Giro remaining the only public Bank in Venice (Soresina 1889, p.8; Dunbar 1892, p. 324; Roberds and Velde 2014, p. 25; Fratianni and Spinelli 2006, p. 271; Ugolini 2017, p. 44). As we shall see, the Banco della Piazza was actually an unintended casualty of the 1629-31 crisis despite not being involved in the monetization of the fiscal response to the shock.

## II

In 1628, the Republic of Venice got involved in the War of the Mantuan Succession.[7] In general, war, famine and epidemics (the so-called "Three Horsemen of the Apocalypse") were often associated in the preindustrial world (Alfani 2013, p. 43), and this episode was no exception to the rule. Starting from March 1629, French and Imperial troops crossed the Alps to participate in the conflict (Alfani and Percoco 2019, p. 1171). In the Italian states, the first decades of the seventeenth century were characterized by severe food shortages (Alfani 2018, p. 152), and the already meagre supply of food was further jeopardized by war. In general, in Venice famine episodes were less intense than elsewhere thanks to the Republic's capacity to collect grain from the rest of the Mediterranean (Todesco 1989, p. 11), but this came at a sizable cost for the government, which sold the grain at subsidized prices to the population (Ugolini 2017, p.37). From a fiscal viewpoint, therefore, the war and famine put considerable pressure on public finances already. As early as April 1629, the governors of mainland cities were complaining to the central government about the severity of the famine and the high costs generated by it.[8]

Starting from spring 1630, an outbreak of bubonic plague (initially transmitted from the North by Imperial troops) started to spread first in the Duchy of Milan and then in the Venetian mainland, finally arriving in the city of Venice in late summer.[9] The massive outbreak

[7] The War of the Mantuan Succession (1628-31) was an Italian episode of the Thirty Years' War (1618-1648) fought on a European scale between supporters and opponents of the Habsburg monarchies. After the death of the heirless Duke of Mantua, two claimants to the succession appeared (the Duke of Guastalla and the Duke of Nevers). The Holy Roman Empire, Spain, and Piedmont supported Guastalla's claim, while France and Venice supported Nevers'. The military outcome being unclear, the conflict was resolved by a political accord (Treaty of Cherasco, 1631), confirming Nevers as Duke of Mantua.

[8] On April 24, 1629, the Podestà (governor) of the mainland city of Bergamo, Valier, wrote to the Venetian Senate, complaining about the fact that the famine was severely hitting his territory (Pederzani 1992, p. 258). According to archival evidence, the price of wheat in Bergamo, which had fluctuated between 40 and 60 liras during the year 1627, reached an extreme height of 140 liras in 1629, to come back to a range between 20 and 60 liras in the years 1631-33 (Archival source: Biblioteca Civica Angelo Mai, Bergamo, 1.2.18.10.5 *Calmieri dei cereali*, 11, class. 1.2.18.10.5-11 (pr. 26781) 1627 gennaio - 1627dicembre; 13,class.1.2.18.10.5-13(pr.26783)1629 gennaio - 1629 dicembre, 14, class. 1.2.18.10.5-14 (pr. 26784) 1631 gennaio-1631dicembre; 15, class. 1.2.18.10.5-15 (pr. 26785) 1632 gennaio - 1632 dicembre; 16, class. 1.2.18.10.5-16 (pr. 26786) 1633 gennaio - 1633 dicembre).

[9] On August 22, 1630, Venetian authorities convened a commission of 36 physicians in order to evaluate the nature and the diffusion of the disease; the majority of the commission (28 members out of 36) were opposed to declaring that the disease was of pestilence nature, notwithstanding plague episodes had already been

in the city occurred between September and December 1630 – 20,923 deaths – with a peak in October 1630 (Ell 1989, p. 130), and in total 43,088 deaths were recorded over just three years; the population of Venice was 141,625 in 1624 and became 102,243 in 1633, a reduction of nearly 30 per cent (Lazzari et al., 2020, p.3). Such figures are consistent with the 35 per cent estimated mortality in Northern Italy, and should be compared to an estimated average annual mortality of between 2.7 and 3.7 per cent in normal times (Lazzari et al. 2020, p. 3). Many different indicators agree that the shock to the domestic economy was colossal, disrupting many diverse aspects of economic life including the arts industry – music production shrank by 40 per cent (Gonzaga Band 2018), while the average price of paintings collapsed by 81 per cent (Etro and Pagani 2010). While traditionally this epidemic has been considered a short-run disaster with limited long-run impact (Rapp, 1976, p. 154), recent research points to the fact that this was a crucial turning point for the Venetian economy (Alfani and Percoco 2019, p. 1197). More specifically, the 1630 plague provided a structural break in the way in which some macro-level variables – population density, urbanization and taxation per capita – affected wealth inequality (Alfani et al. 2020); the plague put the Republic on a lower growth path, favouring the rise of Northern Europe as well as of the Sabaudian State within Northern Italy (Alfani 2020).

To address the pandemic, the Senate had to decide immediately its policy action. During the previous 1576 pandemic the Venetian government had reacted slowly, both denying the plague and downsizing the number and nature of deaths (Palmer 1978, pp. 238-75; Preto 1979, p. 123). The first and most urgent issue consisted of protecting public health by designing and implementing a containment policy. In the Venetian territories urban mortality rates during 1630-1 were severe (ranging from 433 per thousand in Chioggia to 615 per thousand in Verona), while in Venice itself the mortality rate was 330 per thousand, pointing to a certain success of the strict lockdowns implemented there (Alfani and Di Tullio 2019, p. 115). Venice had passed its first legislation to address epidemics in 1423, and a Health Office had been established in 1490 (Palmer 1978, pp. 51 and 85). The Health Office used its authority to close shops, as well as to prohibit auctions and markets (Allerston 1996, p. 279). These measures hit the majority of Venetians, who became unable to work during epidemics (Pullan 1960, p. 26; Biraben 1973, p. 145; Cipolla 1976, p. 42; Allerston 1996, pp. 292-5).

The containment measures "were carried into effect on a colossal scale with full resources of the state" (Palmer 1978, p. 142). The government was aware of such negative signaled in the Venetian mainland (Preto 1979, p. 141). On April 4, 1630, the Bergamo authorities had already noted that an outbreak of plague was in the making (Archival source: Biblioteca Civica Angelo Mai, Bergamo, 60, class. 1.2.3.1-60 1629 dicembre 15 - 1632 maggio 22 *"Consiliorum"* pp. 62-63).

effects on economic activity, so it then tried to alleviate the inhabitants' losses (Pullan 1964, p. 410; Cipolla 1976, p. 41; Biraben 1973, p. 57).[10] In normal times, social expenditures were very low in Venice: for example, available data for 1602 and 1633 – i.e. before and after the pandemic – show that social expenditures were negligible, amounting respectively to 0.2 and 0.4 per cent of total expenditures; in the same years the service of debt amounted respectively to 8.2 and 19.9 per cent (Alfani and Di Tullio 2019, p. 167). But during a pandemic, things were different. In practice, during the 1630 pandemic the government bought necessary goods from merchants to distribute them to the confined population,[11] as it had already done during the 1575 plague (Pullan 1964, p. 409).[12] When districts were put into quarantine, their inhabitants were provisioned by the State (Palmer 1978, p. 143). The overall fiscal effort to help inhabitants included subsidies and other fiscal help given to affected communities, and distributions of free rations of grain (Alfani 2018, p. 162). When decisions to destroy supposedly infected goods were made, compensation was paid – though not always in full (Pullan 1964, pp. 251 and 319; Allerston 1996, pp. 278 and 287). Moreover, the Venetian government influenced employment and nominal wages in the sectors under its total or partial control, including all the activities strictly related to health (esp. body clearers, who got paid huge salaries during pandemics: Allerston 1996, p. 296) and defence (esp. Arsenal workers, who had their wages paid despite being confined to their homes: Pullan 1964, p. 420). These fiscal transfers helped directly or indirectly inhabitants in trouble, particularly those in the lowest classes, which represented the largest part of the overall population: such people were often in debt, and at the greatest risk of crossing the boundary between subsistence and poverty (Alfani and Di Tullio 2019, pp. 62-3). In the absence of subsidization, these people would have been most likely to revolt against the government's containment measures, as the opportunity cost of rioting would have been extremely low to them.

[10] A textile merchant pleaded for the quarantine to be lifted, given that "an incomparable greater number of people has died purely as a result of unemployment than of typhus or any other contagious disease" (Pullan 1964, p. 409); Verona was reported to be suffering more from the lockdown than from the disease itself (Palmer 1978, p. 275). Bribery episodes were registered, merchants being anxious to get their goods into Venice (Palmer 1978, pp. 231-5). Also the regular activity of the Mint suffered during the plague episodes (Stahl 2001, pp. 42-53).

[11] On December 3, 1630, the civil authorities of the city of Bergamo expressed increasing concern about the growing debt that the municipality was accumulating to address the costs of the pandemic (Archival source: Biblioteca Angelo Mai, -60, class. 1.2.3.1-60 1629 dicembre 15 - 1632 maggio 22 "*Consiliorum*" pp. 108-9).

[12] Between December 1628 and November 1629 (i.e., during the early times of the famine), the Venetian Fodder Office ("Provveditori della Biave") had already sent to the mainland territories more than 236,000 ducats worth of grain, a sum which roughly corresponded to 10% of the Republic's total expenditures during the "normal" year 1633 (Pezzolo 2006, p. 84). The authorities of Vicenza and Treviso did not obey letters from Venice that ordered them to send grain to other territories (Lombardini 1963, p.27); similar letters were sent to Bergamo (Archival source: Biblioteca Civica Angelo Mai "260, class. 1.2.2.1-260 1629 dicembre 29; Venezia, "*in nostro ducali palatio", 1629, 29 decembris. Solutio datii traversini per territorium cremense pro bladis*", p. 264) and to Verona (Archivial source: Archivio di Stato di Verona, 34 ,1625 lug. 8 - 1630 mag. 31, "*Registrum Litterarium Ducalium*", p. 171).

## III

The 1629-31 famine and plague thus obliged the Venetian government to implement subsidies on a colossal scale, but how was such a huge fiscal expansion financed? Taxes were actually increased;[13] however, the plague had made tax collection more difficult, increasing the fiscal pressure (Pezzolo 1994, pp. 322-3; Alfani and Di Tullio 2019, p. 29). To help alleviate such pressure, the government also allowed Jews to lend on collateral outside of the ghetto (Preto 1979, p. 144; Allerston 1996, p. 293). But this was far from enough for financing the fiscal expansion. To deal with its worsening deficit, the Republic had to resort to fiscal monetization through the scriptural money issued by the Banco del Giro: in fact, the government paid its creditors by merely crediting their current accounts with the bank. The size of the Giro bank's liabilities, which had generally been less than one million ducats in the 1620s, rose to 2,071,168 ducats in April 1630, and surpassed 2,666,926 ducats in June 1630 (Soresina 1889, pp. 19, 23, and 29; Siboni 1892, p. 290; Roberds and Velde 2014, p. 24). While the Giro bank's liabilities kept increasing in concert with the famine and the bubonic plague (Soresina 1889, p. 29), the asset side of its balance sheet underwent a serious deterioration: the rise in scriptural money held by the public was theoretically matched by an increase in floating (non-marketable) government debt, but in reality such a debt did not actually exist, so that in fact the bank's assets were actually decreasing with respect to its liabilities. This situation, which *de facto* generated a loss for the issuing bank, precisely corresponds to the situation described in Figure 3 above. Fiscal monetization was actually decreasing the net worth of the monetary authority: in view of this, we interpret this episode as an historical illustration of the "net-worth helicopter money" strategy.

The choice of implementing such a radical strategy triggered a number of serious consequences. First, the government-induced growth of the Banco del Giro's business crowded out the activity of the Banco della Piazza di Rialto (Ugolini 2017, p. 44). The balance sheet of the Banco della Piazza had reached its peak of 1.7 million ducats in 1618, i.e. one year before the establishment of the Banco del Giro; in 1630, the amount of deposits with the former had dropped to 56,185 ducats only, making the bank moribund until its final demise in 1637 (Siboni 1892, p. 290; Sissoko 2002, p. 8; Pezzolo 2018, p. 155). Second, the monetary expansion entailed a stark depreciation of the value of Giro bank money, which the bank had no way to counteract. Up to 1625 the premium between Giro bank money and coin had been positive and substantial; then from 1625 the premium began to fall, slowly at first and then precipitously in 1630, eventually turning negative (Soresina 1889, p. 29; Roberds and Velde 2014, p. 24): the *agio* was equal 20 per cent in 1624, then it dropped to 19.5 per cent in early 1629 and fell into negative territory (-10 per cent) in 1630 (Pezzolo 2018, p. 156). The depreciation of Giro bank money can be confirmed by looking at its market value in terms of silver. As shown in Figure 4, during the 1629 famine Giro bank money had already lost 10 per cent of its silver value, and at the peak of the plague (in September 1630) it was down as much as 25 per cent with respect to its pre-crisis valuation.

[13] On August 4, 1629 a new lump-sum wealth tax was introduced, and a second one was added in 1630 (Preto 1979, p. 144; Pezzolo 2003a, p.69; Pezzolo 2003b, p. 110; Pezzolo 2021, p. 74). Moreover, the "duties" – i.e. the set of taxes that affected the transit and consumption of goods (Alfani and Di Tullio 2019, p. 24) were increased twice in 1629, and once more in 1630 (Pezzolo 2021, p. 71).

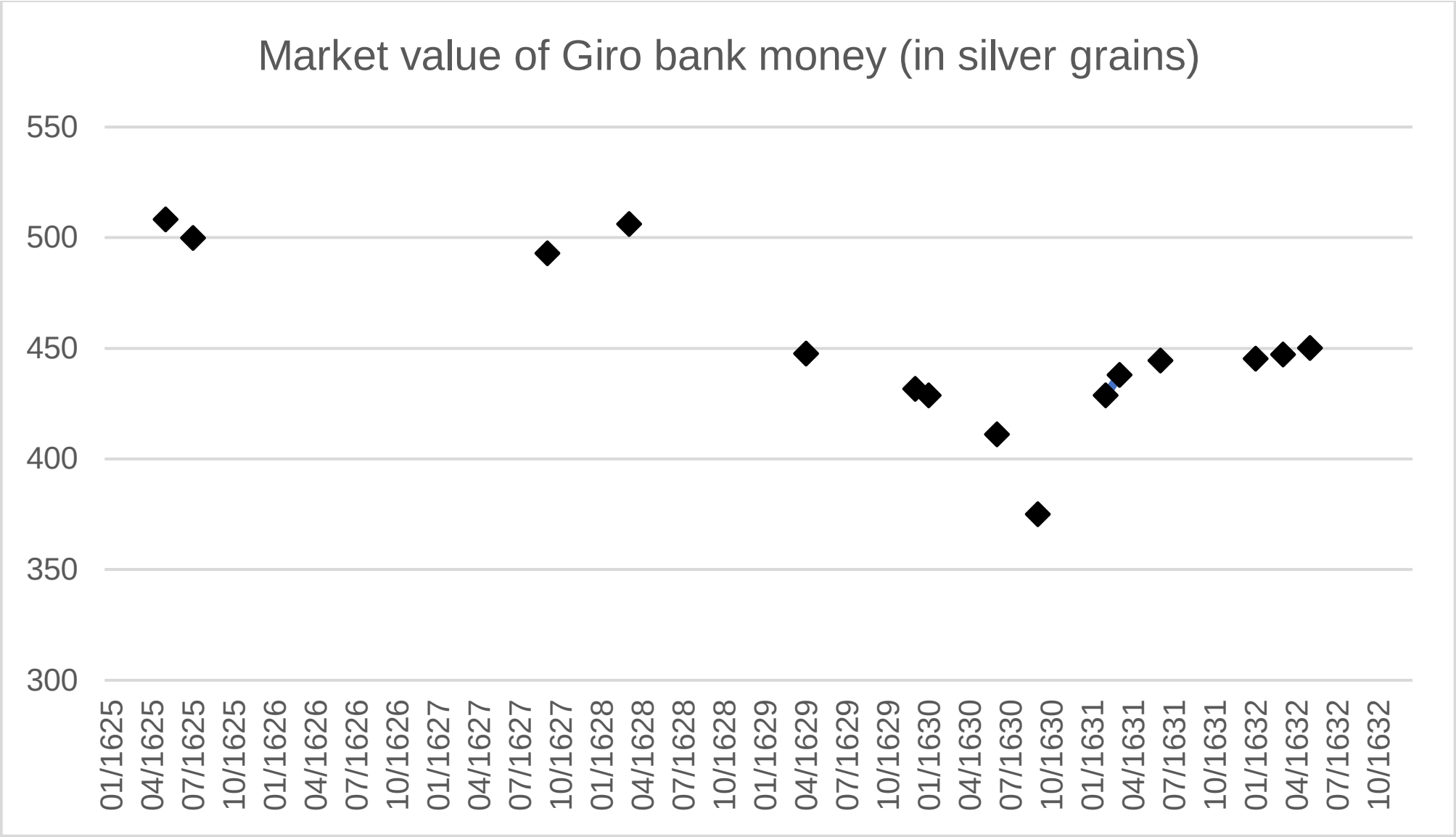

*Figure 4: Market value of Giro bank money (in silver grains). Source: Mandich (1957, p. 1173).*

The overexpansion of the money supply thus quickly triggered a substantial monetary depreciation, which the bank had no means to counteract in view of the weakness of its balance sheet. This depreciation forced the government to backtrack on its “net-worth helicopter money” strategy and reform its monetary policy setting. Already in July 1630, a monetary committee (*Inquisitori del Banco Giro*) was established, with the aim of reducing the bank's liabilities (Soresina 1889, p. 23; Siboni 1892, p. 290; Roberds and Velde 2014, p. 24). The reform proposed by the committee was made effective on September 24, 1630. Under this plan, the accounts of a number of separate administrations and public concessionaries – worth 716,652 ducats – were removed from the Giro Bank and transferred to the Mint, which was the division of the Treasury which was charged with the management of the public debt (Pezzolo 2003a). Private depositors were moreover invited to convert their Giro bank liabilities into "Mint deposits", paying seven per cent interest (Soresina 1889, pp. 25-6; Siboni 1892, p. 290, Roberds and Velde 2014, pp. 24-5). As the so-called "Mint deposits" were not current account deposits but inscribed bonds (Pezzolo 2003a), this amounted to the conversion of unremunerated sight liabilities into interest-bearing long-term bonds. Furthermore, Mint revenues from sales of life annuities at 14% were applied to the Giro bank (Soresina 1889, p. 26; Roberds and Velde 2014, p. 25) to strengthen its financial position. In accounting terms, this means that the Treasury first increased the outstanding government debt in order to raise the liquidity that allowed the bank to withdraw some of the money it had issued (Figure 5.1), then it gratuitously provided it with the resources transferred to it: in so doing, one of the Bank's liabilities was reduced without any asset being diminished, which *de facto* entailed an increase in the Bank's net worth (Figure 5.2). All in all, this amounted to transferring the loss previously generated on the monetary authority's balance sheet (Figure 3.2) back to the fiscal authority's balance sheet (Figure 5.2): differently said, the September 1630 reforms *de facto* reversed the "net-worth helicopter money" strategy implemented since the beginning of the macroeconomic shock.

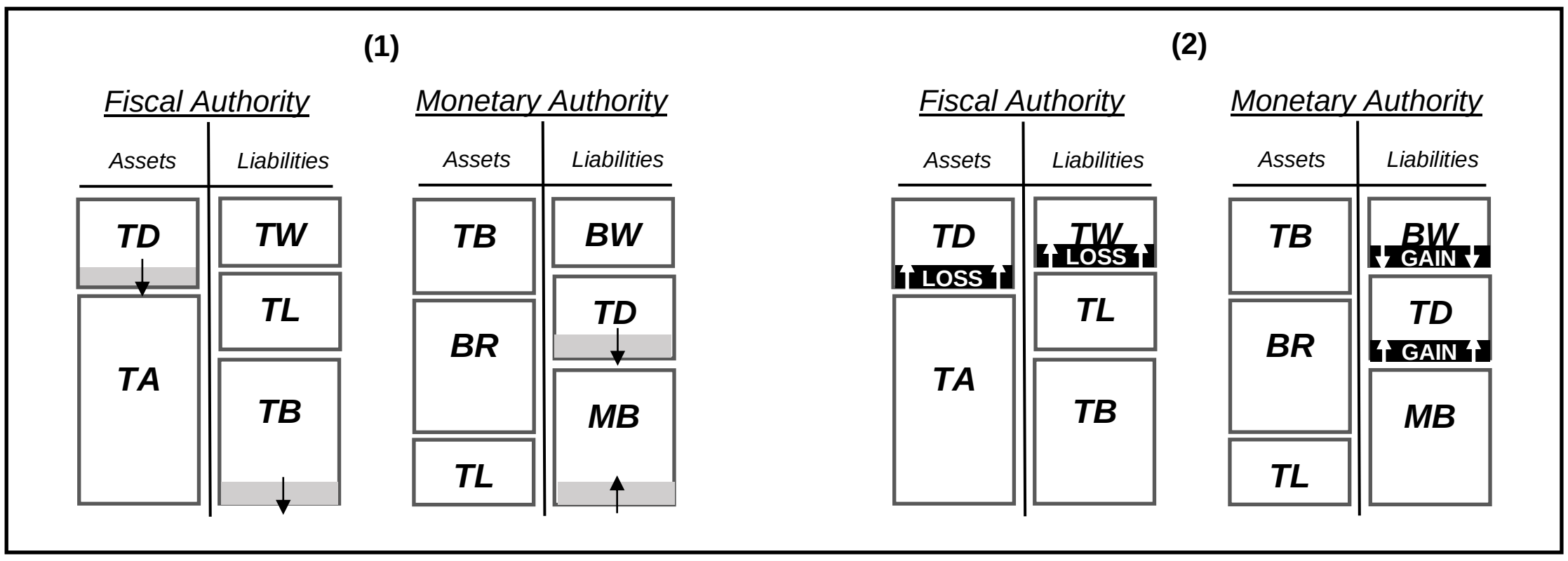

*Figure 5. Balance sheet effects of the September 1630 reform.*

These combined operations reduced the Giro balance sheet and then the money supply, thus allowing for a reappreciation of the bank money (see Figure 4) – although at the price of an increase in the funded public debt. The new monetary policy strategy brought down Giro balances to 1.4 million ducats at the end of 1630 (Soresina 1889, p. 29; Roberds and Velde 2014, p. 25), but it was not sufficient to restore convertibility on demand of Giro bank liabilities (Dunbar 1892, pp. 327 and 330). Moreover, the government was not able to stabilize the Giro balance sheet below 900,000 ducats until 1638 (Roberds and Velde 2014, p. 25), which from then on represented the usual amount of the Giro bank balance sheet (Pezzolo 2021, p. 96). Ex post, price instability – "an unprecedented rise in prices has been the worst blow of all", as argued by a contemporary observer (Cohen 1988, pp. 134-5) – and currency devaluation (Roberds and Velde 2014, p. 23) had been the macroeconomic outcomes of the "net-worth helicopter money" strategy. So, it is unlikely that such a strategy was socially optimal, especially in view of the fact that the relatively severe indebtedness of the Giro bank was not particularly dramatic compared with the Republic's fiscal income (Pezzolo 2003a, p. 64).[14] Meanwhile fiscal dominance was in place, with the Venice Senate completely controlling the Giro bank governance (Anonimo 1847, p. 364; Soresina 1889, p. 8; Dunbar 1892, pp. 312 and 321). So why did the Venetian authorities choose to implement such a suboptimal strategy – only to quickly backtrack from it? To answer this question, it is natural to examine possible links between fiscal monetization and political pressure.

In the Republic of Venice, wealth inequalities were extreme, both between the lower and middle class and between the middle and upper class (Alfani and Di Tullio 2019, pp. 105-12). This reflected the distribution of political rights, which was strictly reserved to the upper class. However, when calamities occurred, the patricians had to take into account the expectations of the lower classes. Urban populations were watchful of rulers, and they were ready to riot if they became convinced that the government was not doing all it could, (and should) have done to ensure the availability of food, guaranteeing the "right to bread" (Alfani 2018, p. 162). Politicians had much to fear, also in terms of personal safety, from riots motivated by distributional reasons – "injustice" – so that incentives to act were really strong. Indeed, most popular riots in early modern Venice seem to have been caused by "political" claims. Inhabitants' preferences mattered, although even full citizenship did not guarantee full political rights, as those were reserved for the patriciate (Alfani and Di Tullio 2019, pp. 13 and 61).

[14] The Republic's fiscal income was worth 3,861,827 ducats in 1621 and 2,949,888 in 1633 (Besta 1912, pp. 464-75 and 486-93).

The extraordinary money-financed fiscal stimulus implemented in Venice during 1629-30 can be considered optimal from the perspective of poor inhabitants, who benefited from real subsidies without being sensitive to the "monetary externalities" generated by the policy.[15] In this respect, the government decision can be considered consistent with the aim of pleasing the majority of inhabitants during a big macroeconomic downturn, thus enhancing consensus and avoiding riots. The theoretical framework behind this interpretation is illustrated in the Appendix. At one point, however, monetary instability became too strong and the government risked losing control of the useful monetization mechanism. At that point, those that were mostly sensitive to "monetary externalities" (i.e. the ruling upper class) were ready to accept some "sacrifices" in order to put an end to currency depreciation. The "sacrifices" consisted of accepting a conversion of liquid sight debt (Giro bank money) into illiquid long-term debt (the "Mint deposits"). Acceptance of this conversion (somewhat akin to a debt restructuring) by patricians allowed for reversing part of the monetization and putting an end to instability. Importantly, the "sacrifices" were such only in the short term, as in the long term high interest rates were duly paid to the holders of the newly-created funded debt. Because the conversion was actually financed through earmarking future tax revenues, and because the tax system was strongly regressive in Venice, the short-term redistribution in favour of the poor was offset by a long-term redistribution in favour of the wealthy. Indeed, the 1629-30 money-financed fiscal expansion did not cause a permanent change in the condition of poor inhabitants, since analysis of income distributions shows that, contrary to the Black Death of 1348, the bubonic plague of 1630 did not trigger a phase of sustained inequality decline (Alfani and Di Tullio 2019, p. 116; Alfani 2020, p. 204).

In view of all this, we can tentatively provide some speculative interpretations why a "net-worth helicopter money" strategy was first implemented and then abruptly reversed in Venice during the 1629-31 shock. At the beginning of the shock, the ruling class (which held the overwhelming share of Venetian public debt) was apparently unwilling to see the funded debt increase, thus forcing the Treasury to monetize the politically inevitable stimulus. After realizing the substantial costs of excessive monetization, however, consensus was found among patricians on the need for the funded debt to grow in order to limit money depreciation. Contrary to the "monetary-base helicopter money" strategy, the "net-worth helicopter money" might have prevented the long-term government debt from growing in the short term, but the latter's impact on monetary stability was probably much more violent than the former's. As a result, the "net-worth helicopter money" strategy had to be quickly reversed and replaced by a much more conservative policy.

[15] For a definition of "monetary externality", see the Appendix.

## IV

Recently, monetary policy theory highlighted the relevance of the way interventions are orchestrated between the central bank's and the government's balance sheets (Sims 2004; Reis 2015; Orphanides 2016; Benigno and Nisticò 2020). In particular, radical options like the so-called "net-worth helicopter money" have been discussed as a possible strategy to cope with extraordinary macroeconomic shocks such as pandemics. However, episodes in which the "net-worth helicopter money" strategy has been actually implemented appear to be very rare in history. In this article, we have focused on one historical example in which a pandemic recession was actually addressed through the implementation of this strategy.

In the Republic of Venice, the 1629 famine and the 1630-1 plague caused a unique negative macroeconomic shock, that the oligarchic government addressed using a particularly radical form of fiscal monetization that corresponds to the modern notion of "net-worth helicopter money". As a matter of fact, the expansion of the bank of issue's liabilities was associated with a deterioration in the quality of its assets, thus producing capital losses to the issuer and reducing its net worth. This policy entailed monetary depreciation and instability, so that the government had to reverse it very quickly – i.e. before the end of the macroeconomic shock. Backtracking on the "net-worth helicopter money" strategy implied transferring the losses suffered on the monetary authority's balance sheet to the fiscal authority's balance sheet: this actually took place through a de facto government bailout of the bank of issue, as the bank's sight liabilities were converted into long-term public debt.

Why did the government choose to implement this suboptimal strategy? We argue that they did so to avoid political disturbances and popular rioting, while also preventing the long-term government debt from growing. Thus, in the short term, potential political and social consequences of the macroeconomic shock were minimized by implementing a redistributive policy from the rich to the poor. But in order to keep the monetization mechanism viable and prevent a complete debasement of bank money, the quality of the assets side of its balance sheet had to be restored: losses had to be covered through a *de facto* bailout, which made the inevitable growth of long-term government debt eventually occur.

However, it is worth underlining that in the long term, the conversion of sight bank liabilities into long-term debt actually reversed the distributional effects triggered by the "net-worth helicopter money" strategy: as taxes were strongly regressive in Venice, the ensuing conservative fiscal policy which allowed repaying the debt entailed a redistribution of resources from the poor to the rich.

The history of the Venetian reaction to the 1629-31 famine and pandemic echoes many aspects of the Covid-19 crisis. For one thing, it proves that nowadays' extraordinary fiscal expansion to cope with a pandemic were far from unprecedented. Moreover, it suggests that nowadays' central bankers' refusal to embark into some "net-worth helicopter money" experiment may have been a good idea after all. More generally, the Venetian experience with "net-worth helicopter money" highlights the redistributive implications of the design of macroeconomic policies, as well as the importance of political economy factors in the choices underlining such designs.

## ***Appendix: Pandemic Recession, Helicopter Money and Political Pressure***

What happened in Venice in the period 1629-1631 can be also described using a model, using and modifying the theoretical settings introduced in Masciandaro and Passarelli (2019), Masciandaro (2020) and Favaretto and Masciandaro (2021).

### *1. The Policymaker*

The economy consists of a population of inhabitants, and a policymaker that controls both fiscal and monetary policies. In the case of Venice, the Senate represents the incumbent policymaker. Initially we use the standard assumption that the policymaker is a benevolent player (Acemoglou et al. 2020; Argente et al. 2020; Brotherhood et al. 2020; Glover et al. 2020). This assumption will be modified later. All players are rational and share the same information; they maximize their utility in a simultaneously and one-shot way. Moreover, the population size is one, then total and per-capita amounts are the same for all the variables.

Since the population size is one, maximizing the utility of the average representative inhabitant amounts to maximizing social welfare. Without macroeconomic shocks, fiscal monetization is not needed. In Venice in normal times the Republic produced systematic fiscal surpluses. If a pandemic and/or a famine occurs, the policymaker sets her economic policies to maximize social welfare. Venice was hit by an adverse macroeconomic shock from April 1629 – when a famine started – to September 1631 – when a bubonic plague ended.

A pandemic shock triggers a special recession, because inhabitants' incomes can be hit in an heterogeneous way for three reasons. First, the effects of both the disease and the corresponding public policies are unequally distributed (Glover 2020; Bloom et al. 2021). Second, the less the policymaker is involved in supporting the economy during the pandemic recession through fiscal transfers, the more likely are negative second-round effects on the well-functioning of the economic and financial system after the pandemic (pandemic externalities) (Acharya and Steffen 2020; Anderson et al. 2020; Deb et al. 2020; Bloom et al. 2021). Third, the fiscal policy financing introduces the possibility of monetary stability risks (monetary externalities). Therefore the policymaker will choice its economy policy design maximizing a welfare function with three terms (Masciandaro and Passarelli 2019; Masciandaro 2020):

$$V(\beta,\delta,\tau) = U(\beta,\delta,\tau) - F\ (\beta) - M(\beta,\delta) \qquad (1)$$

Where U(β,δ,τ), F(β) and M(β,δ) are respectively the inhabitants' utility, the pandemic externalities and the monetary externalities, while τ, β, and δ represent the key economy policy variables: taxation, fiscal spending and fiscal monetization. Then the first step is to explore how inhabitants behave.

*2. The Inhabitants*

We assume that inhabitants are risk neutral, and they maximize their overall utility from consumption and disutility from effort. In our setting inhabitants' utility is associated with heterogeneous resources (Krishnamurthy and Vissing-Jorgensen 2012; Masciandaro and Passarelli 2019; Glover et al. 2020; Gertler 2020; Masciandaro 2020; Reis 2020), and these sources are combined to consume a single final good (Masciandaro and Passarelli 2019; Glove et al. 2020).

Inhabitants expect that when a pandemic occurs their incomes can be hit. Containment measures save lives, but in parallel impose limitations on several economic activities. People suffer because lockdown measures and quarantines reduce their incomes and expenditures (Baker et al. 2020a; Carvalho et al. 2020; Cox et al. 2020). At the same time, inhabitants expect that the policymaker will help those of them who are suffering with an injection of a lump-sum fiscal transfer to mitigate the pandemic costs (Acemoglou et al. 2020; Argente et al. 2020; Brotherhood et al. 2020; Glover et al. 2020).

As much as inhabitants' losses due to the pandemic can be heterogenous, the same will be true for the distribution of the fiscal transfers (Bayer et al. 2020; Glover et al. 2020). The pandemic shock and the consequent fiscal transfer policy influence inhabitants' welfare in an unequal way, producing a special case of income heterogeneity (Auerbach et al 2020; Bayer et al. 2020; Gertler et al. 2020; Glover et al. 2020; Kaplan et al. 2020).

Given the risk that an activity is frozen during a pandemic due to the containment policies – pandemic or "quarantine shock" (Bayer et al. 2020) – we can distinguish between safe activities and risky activities. Occupations that do not suffer losses during a pandemic produce safe incomes, while risky earnings are associated with activities that are negatively influenced when a pandemic happens (De Vito and Gomez 2020; Elenev et al. 2020). Safe incomes can be taxed. As extreme examples of the two situations during the 1630 pandemic, consider the case of the second-hand clothiers as inhabitants involved in risky activities, being systematically hit by blockades and closures, were the Arsenal workers were individuals with a safe occupation, having their salaries in any case, due to the Senate subsidies.

Therefore the representative unhabitant's utility is:

$$l(1-\tau) - U(l) + \theta(r) \tag{2}$$

Assuming a normalized productivity to one, the first term is the after-tax income, while the second term is a standard increasing and convex effort function. Given taxation, the inhabitant chooses her optimal effort. The optimal condition yields inhabitants' labour supply, decreasing in the taxation, and this rate represents her elasticity to tax distortion. Since both population size and productivity are equal to one, the labour supply represents the total safe income.

The third term represents the risky earning utility. There is only one risky activity, and it is measured using the variable *r*, which parameterizes the risk that the inhabitants bear. The return of the risky activity is θ(r), with θ'(r)>0 and θ''(r)<0. We normalize this return to one. If a pandemic occurs, with probability *p*, lockdown and quarantine are implemented. The inhabitants know that the value of the risky activity will fall – for simplicity to zero – and they will bear the full cost. At the same time, the inhabitants expect that the policymaker will design a fiscal transfer policy to address their losses, in a proportion β of such losses, and with a corresponding monetization δ.

Safe incomes and risky earnings finance consumption. The overall budget constraint of the representative average inhabitant will be:

$$c = l(\beta, \delta, \tau) + \theta(\beta, \delta) \tag{3}$$

With risky earnings being the only source of heterogeneity among inhabitants, such a metric allows us to highlight in the clearest and simplest way the redistributive effects of the fiscal transfer and its monetary financing. Given that the economy policy design influences the inhabitants' welfare, now we turn again our attention to the policymaker choices.

*3. The Economic Policy Design*

When a pandemic breaks out, the policymaker designs a containment policy, facing an unpleasant dilemma between two public goals (Baldwin and Weder di Mauro 2020). The policymaker needs to protect public health by implementing a containment policy with the aim of minimizing the expected loss of life. In parallel, containment policies save lives, but, given the interactions between economic decisions and epidemics (Eichenbaum et al. 2020), any containment policy has short-term economic and financial costs (Deb et al. 2020; Ludvigson et al. 2020). These costs simultaneously affect aggregate supply (Del Rio-Chanona et al. 2020; Koren and Peto 2020) and aggregate demand (Andersen et al. 2020; Del Rio-Chanona et al. 2020).

The policymaker can address the pandemic recession by implementing an extraordinary fiscal transfer policy using a lump-sum distribution (Bloom et al. 2021; Glover et al. 2020), with the aim of mitigating the negative effects of containment measures (Beck 2020; Bénassy-Quéré et al. 2020; Brunnermeier et al. 2020; Deb et al. 2020; Drechsel and Kalemli-Ozcan 2020; Gros 2020a; Kahn and Wagner 2020; Segura and Villacorta 2020). The fiscal transfer can come in many forms, as income subsidies, work insurance, equity injections, loan guarantees (Céspedes et al. 2020; Didier et al. 2020; Elenev et al. 2020). Transfer payments can be unconditional (Kubota et al. 2020) and/or conditional on the inhabitant's status (unemployed and/or liquidity or credit constrained individuals) (Bayer et al. 2020). In Venice the Senate systematically implemented fiscal transfer policies during the pandemic recessions. Then two questions arise.

First, how large should this fiscal policy be? Two opposite options arise. At one extreme, the policymaker is completely absent, and inhabitants suffered income losses. At the other extreme, the fiscal expansion helps the suffering inhabitants. The policymaker injects resources in the economy, and the metric of this fiscal action is a proportion, $\beta \in (0,1)$ of the inhabitants' losses, which is the policy variable that parameterizes fiscal policy.

Second, how can such a fiscal policy be financed? The policymaker can raise taxation (Bloom et al. 2021; Eichenbaum et al. 2020), issue debt or money. The policymaker defines the optimal fiscal transfer policy, $\beta^*$, and this policy can be financed by issuing new debt, charging a regressive lump-sum tax $\tau$ on the safe income for servicing the issued debt, and through monetization. Assuming no default risk, the policymaker budget constraint is:

$$\beta\theta\big(1 + i(1-\delta)\big) = \tau l \qquad (4)$$

where $\tau$ is the lump-sum tax, $l$ is the safe income of the inhabitants before taxes, $i$ is the interest paid on the public bond and $\delta$ is the fiscal monetization where $\delta \in [0,1]$. For any unit of debt issued, the policymaker repays $1+i(1-\delta)$. The cost of debt, $i(1-\delta)$, is negatively associated with the degree of fiscal monetization. When a higher monetization is implemented (i.e. higher $\delta$), a lower portion of funded debt will be sold to inhabitants.

Therefore fiscal transfer and its monetization influence inhabitants' consumption. But the pandemic policy can produce long standing effects on inhabitants' welfare. So we introduce the possibility of monetary and pandemic externalities, internalizing future negative spillovers due to the economic policy action that can affect the economy when the pandemic ends.

First, fiscal monetization is not a free lunch: it may create monetary externalities. Monetary externalities can depend on the association between central bank seigniorage and monetary stability risks. The more traditional channel is the relationship between seigniorage and inflation tax (Buiter 2007), that increases both national inflation (Friedman 1969; Aizenman 1992) and, via exchange rate devaluation, international inflation (Hamada 1976). Moreover, monetary externalities can also include banking (Bianchi, 2010) and financial (Stein 2012; Cesa Bianchi and Rebucci 2017) imbalances, or more generally it is a device to take into account the risk of monetary policy multiple equilibria and their costs (Gliksberg 2009; Airaudo and Bossi 2017).

Fiscal monetization threatens the monetary stability goal in the post pandemic period, as it has been the case during the 1630 pandemic recession. The costs of monetary instability, $M = M(\beta, \delta)$, are quadratic in the degree of monetary accommodation $\delta$:

$$\frac{\Phi}{2}\delta^2\beta\theta \equiv M(\beta, \delta) \qquad (5)$$

The monetary externality aversion – i.e. the parameter $\Phi$ - is homogenous among inhabitants. With this assumption it will be evident that it is sufficient to have just two sources of heterogeneity among inhabitants – fiscal transfer and its bond financing – to have a multiple equilibria setting in terms of political consensus. With further heterogeneity sources the results should be even stronger.

Second, the less the policymaker is involved in supporting the economy, the more likely are negative second-round effects on the well-functioning of the economic and financial system after the pandemic (Acharya and Steffen 2020; Anderson et al. 2020; Deb et al. 2020). The absence of active public policies can have adverse economic effects that spread out over time and into the longer term, as a reduction in return to human capital or negative structural changes in terms of trading patterns and stalling development (Bloom et al. 2021). On this respect, and notwithstanding the public action, the 1630 pandemic recession was a crucial negative turning point in the history of the Most Serene Republic.

To capture in the simplest way this channel, let the pandemic externality function be:

$$\frac{\varepsilon}{2}[(1-\beta)\theta]^2 \equiv F(\beta) \qquad (6)$$

The pandemic externalities are increasing and convex in the amount of losses, and they are lower the higher are the fiscal transfers, $\beta$. Also the pandemic externality aversion – the parameter ε - is homogenous among inhabitants, for the same motivations above expressed.

### *4. The Optimal Helicopter Money*

Inhabitants and policymaker simultaneously optimize their choices. The average representative inhabitant, optimizing the goal function (2), and given her elasticity to tax distortion, $\eta$, identifies the optimal effort, $l^*$ and the optimal risk assumption, $\vartheta^*$. The corresponding safe incomes and risk earnings finance consumption. This assumption is particularly relevant during a pandemic: shutdowns and quarantines lockdowns produce material deprivation and households can draw on all their net available resources to address the shock (Baker et al. 2020a; Carvalho et al. 2020; Cox et al. 2020). Moreover the available resources depend on the design of taxation, fiscal transfer and income from bond financing. Equation (3) becomes:

$$c = l^*\big(1-\tau(\beta,\delta)\big) + \beta\theta^*\big(1+i(1-\delta)\big) \equiv C(\beta,\delta) \qquad (3')$$

The policymaker maximizes the social-welfare function (1), setting her strategy on taxation, $\tau$, fiscal transfer, $\beta^*$, and fiscal monetization policy, $\delta^*$. Being a social planner in action, fiscal and monetary policy are optimally coordinated (among others, from Abel 1987 to Bianchi et al. 2020), including the degree of fiscal monetization (among others, Chari and Kehoe 1999, Punzo and Rossi 2019). Regarding the institutional setting, we have here a fiscal dominance regime (Sargent and Wallace 1981): monetary policy is not independent. Focusing on the optimal level of fiscal monetization, $\delta^*$, its social optimal value is:

$$\delta^* = \frac{\eta}{1-\eta}\frac{i}{\phi}. \qquad (7)$$

The optimal level of monetization, $\delta *$, will increase: a) the more taxation is distortionary; b) the more the cost of debt servicing is high; c) the more monetary externality aversion is low.

Given the decision in terms of fiscal monetization, the final step is its implementation. Here a central bank - or a state bank issuer, the Venice Giro Bank, in the case of Venice – comes in as a public institution with its goal, which comes from somewhere or someone (Reis 2013) – the Venice Senate. The central bank, taking into account its resource constraint, $\xi$, technically implements the policy choice, using case by case the more effective tool (Castillo Martinez and Reis 2021). Focusing on helicopter money policies, two options are available (Galì 2020a; Benigno and Nisticò 2020): changes in central bank liabilities (soft helicopter money) and/or changes in the central bank net-worth (hard helicopter money). The overall macroeconomic effects of these policies are disputed (Bernanke 2003 and 2016; Woodford 2012; Turner 2013; Perotti 2014; Muellbauer 2014; Borio et al. 2016; Di Giorgio and Traficante 2018; Bartsh et al. 2019; Galì 2020a; Benigno and Nisticò 2020; Bartsh et al. 2020). For our purposes it is sufficient to assume that the central bank defines its optimal helicopter money action, discounting its effects on monetary externalities:

$$\xi = f(\delta^{*}) \tag{8}$$

## *5. Political Pressure and Helicopter Money*

Now, we can see what happens if the policymaker is not the benevolent policymaker. If politicians are in charge and at the same time inhabitants are heterogenous, different monetization policies have associated redistributive effects, and at the same time such policies can have political effects if the political consensus depends on inhabitants' economic preferences (Masciandaro and Passarelli 2019; Masciandaro 2020; Favaretto and Masciandaro 2021).

The net transfers implied by social optimal policies can be positive for some and negative for others. Moreover, if a policy task has distributional effects, politicians would like to control those effects (Alesina and Tabellini 2007): the redistributive effects are relevant as long as the politicians care about the inhabitants' preferences. For example, one way to build consensus in favour of containment policies is to use fiscal retributive policies to reduce the costs to those whose resources are threatened by shutdowns and quarantines (Glover 2020). Therefore, we need to explore the inhabitants' preferences regarding the policy mix designed by the policymaker. Two different dimensions are relevant: inhabitants can be or not subsidy recipient individuals, and/or they can be or not monetization prone agents.

Which are the subsidized inhabitants? Let us consider any inhabitant $j$, being $\theta + \theta^j$ the amount of risky earnings in her balance sheet. With $\theta^j > 0$ inhabitant *j* will be a subsidized inhabitant relative to the average inhabitant (subsidization gain) . Let $L(\theta^j)$ be the distribution of the subsidized inhabitants across the population. With risky earnings being a proxy for the fiscal transfer, these resources in the balance sheet of the median inhabitant tell us whether the subsidized inhabitants represent the majority or a minority of the population.

What about monetization propension? Inhabitants can be heterogeneous also as funded debt holders. In this case, the more a inhabitant *j* is a debt holder, the more she will be monetization adverse, given that more monetization implies lower interest rates. Let $(\beta + b^j)(1-\delta)\theta$ be the amount of bonds in inhabitant *j*'s balance sheet. With $b^j < 0$ inhabitant *j* will be a monetization prone individual relative to the average inhabitant (monetization gain). The bond holding of the median inhabitant signals whether the monetization prone inhabitants represent the majority or a minority of the population. Therefore, given the general individual utility function (1) and the above definitions of $\theta^j$ and $b^j$, the inhabitant *j*'s utility $V^j(\beta,\delta)$ is:

$$V^j(\beta,\delta) = V(\beta,\delta) + \beta\theta^j + b^j\theta i(1-\delta) \qquad (9)$$

where the last two terms on the right-hand side account for the two forms of heterogeneity of individual *j* relative to the average inhabitant. Each inhabitant's preferences can differ from those of the benevolent policymaker because of these two terms. Focusing on monetization preferences, the optimal fiscal monetization for inhabitant *j* is:

$$V_\delta^j = V_\delta - b^j\theta i \leq 0 \qquad (10)$$

Assuming equation (10) holds as an equality, solving it yields:

$$\delta^j = \left(\frac{\eta}{1-\eta} - \frac{b^j}{\beta}\right)\frac{i}{\phi}. \qquad (11)$$

By comparing equation (7) with the social optimal monetization (11), it is evident that, given a fiscal policy $\beta \neq 0$, a political distortion can arise between a inhabitant's preferred policy and the social optimal monetization:

$$\delta^j - \delta *= -\frac{b^j}{\beta}\frac{i}{\Phi} \qquad (12)$$

The political distortion $\left|\hat{\delta} - \delta^*\right|$ will reflect inhabitants' preferences. The direction of the political pressure depends on who the median inhabitant is. For example, we can assume that financial asset/wealth holdings are very skewed, concentrated among a small segment, the rich, of the population. Therefore monetary policy, influencing asset returns, produces redistributive effects that benefit the holders of such as assets (Krishnamurthy and Vissing-Jorgensen 2011; Brunnemeier and Sannikov 2014).

We interpreted the monetary stance in Venice in early modern times as an extraordinary "hard helicopter money" with redistributive effects. Then a question arises: which inhabitants like fiscal monetization? Among all the possible equilibria, all Venetian subsidized inhabitants like helicopter money, but when they were monetization adverse individuals. In parallel, all monetization prone inhabitants like helicopter money, but when they are not subsidized inhabitants. In other words, uncertainty is present in cases without a clear-cut net benefit, as subsidized merchants that are also bond holders.

But how relevant are the median inhabitant's preferences for the incumbent policymaker? Taking inspiration from Passarelli and Tabellini (2017) and Favaretto and Masciandaro (2021), we assume that the monetary policy decisions are associated with political consensus, because consensus depends on the median inhabitant's preferences through economic and psychological group-thinking mechanisms. The risks of political unrest can influence incumbent policymakers, and these risks can be motivated by facts and emotions. If the policymaker considers the median inhabitant's preferences as a relevant proxy for riot risks, political pressures may be relevant in shaping fiscal monetization choices. The link between inhabitants' preferences, political pressure and political choices can emerge also in an oligarchy of merchants, as the Venice Republic was at that time. In fact, ordinary Venetians used collective actions to influence patrician choices, especially during crisis periods.

This situation is captured in the simplest way assuming that the actual monetary policy decision $\delta_A$ is such that:

$$\delta_A = \chi \left|\delta^j - \delta^*\right| \qquad (13)$$

where $0 < \chi < 1$ represents the relevance of the political pressure. In Venice, the 1630 extraordinary monetization over-expansion created inflation and currency depreciation. In parallel, the population expected "whatever it takes" myopic fiscal policies, and the politicians tended to please inhabitants' preferences, given the threat of riots. Political pressure and helicopter money were likely to be two sides of the same coin.

*6. Further Steps*

The analysis can be enriched in several directions:

a) Monetary externality sensibility and inhabitants' heterogeneity: monetary instability is assumed to be an homogeneous social cost. But inhabitants can be heterogeneous in their ability to address such risks through hedging, with some individuals facing – or feeling that they face – higher costs due to monetary instability (i.e. *inflation-adverse* citizens). In other words, we could explicitly take into account the redistributive effect of inflation, that have long been recognized in the traditional literature (Keynes 1923; Bresciani-Turroni 1937; Friedman and Schwartz 1963), and that has been discussed again recently (Doepke and Schneider 2006; Colbion et al. 2012). Also in early modern Venice the ruling elites constantly demanded a stable currency (Al-Bawwab 2021). Allowing for this kind of heterogeneity would lead to a straightforward prediction: the smaller the mass of inflation risk-adverse citizens, the stronger the political pressure to engage in fiscal monetization.

Moreover, we could introduce heterogeneity in the propensity to consume, that can influence the effect of the fiscal transfer in stimulating consumption (Andreolli and Surico 2021). This change could help to explain the empirical estimates of the marginal consumption propensity during a pandemic (Baker et al. 2020b; Chetty et al. 2020; Coibion et al. 2020; Karger and Rajan 2020; Kim and Lee 2020; Kubota et al. 2020). Finally, we can assume heterogeneity in inhabitants' marginal propensity to take risk (Kekre and Lenel 2021), and also this heterogeneity can influence the distribution of the fiscal transfer effect.

b) Taxation and inhabitants' heterogeneity: safe income taxation has been assumed to be the same for all individuals. In the presence of taxation heterogeneity, the distributional effects are likely to increase. For example, given the decisions regarding the fiscal policy and its monetization, if richer inhabitants are likely to have higher tax burden, all else equal, they would prefer lighter fiscal policies. The income and/or taxation heterogeneity can be relevant in strengthening or weakening political pressure in favour or against the fiscal monetization.

c) Public debt and interest rates: public debt is only issued to address the pandemic-related recession and the interest-rate level remained constant. Assuming an initial debt, or interest rate endogeneity depending on the debt stock, would exacerbate the policy trade-offs and, consequently, the relevance of the political distortions.

*Appendix: References*